\documentclass[usenatbib]{mn2e}
\usepackage{times,txfonts,graphicx,natbib,tabularx, amsmath1, aas_macros}

\usepackage[dvipsnames]{xcolor}
\usepackage{upgreek}
\usepackage[normalem]{ulem}
\usepackage{float}
\usepackage{hyperref}
\newcommand{\be}{\begin{eqnarray}}
\newcommand{\en}{\end{eqnarray}}
\newcommand{\pa}{\partial}
\newcommand{\f}{\frac}

\newcommand\bb[1]{\mbox{\boldmath{$#1$}}}
 
\newcommand\bcdot{\bb{\cdot}}
\newcommand\btimes{\bb{\times}}

\title[Modified MTI]{Effect of Composition Gradient on Magnetothermal Instability Modified by Shear and Rotation}
\author[Gupta et al.]
{Himanshu Gupta$^{1}$\thanks{Email: hiugupta@iitk.ac.in}, Anya Chaudhuri $^{1}$\thanks{Email: chaudhuri.anya@gmail.com }, Shubhadeep Sadhukhan$^{1}$\thanks{Email: deep@iitk.ac.in},  and Sagar Chakraborty$^{1}\thanks{Email:sagarc@iitk.ac.in}$
\\
$^{1}$ Department of Physics, Indian Institute of Technology Kanpur, U.P.-208016, India.}

\begin{document}

\maketitle

\label{firstpage}

\begin{abstract}
We model the intracluster medium as a weakly collisional plasma that is a binary mixture of the hydrogen and the helium ions, along with free electrons. When, owing to the helium sedimentation, the gradient of the mean molecular weight (or equivalently, composition or helium ions' concentration) of the plasma is not negligible, it can have appreciable influence on the stability criteria of the thermal convective instabilities, e.g., the heat-flux-buoyancy instability and the magnetothermal instability (MTI). These instabilities are consequences of the anisotropic heat conduction occurring preferentially along the magnetic field lines. In this paper, without ignoring the magnetic tension, we first present the mathematical criterion for the onset of composition gradient modified MTI. Subsequently, we relax the commonly adopted equilibrium state in which the plasma is at rest, and assume that the plasma is in a sheared state which may be due to differential rotation. We discuss how the concentration gradient affects the coupling between the Kelvin--Helmholtz instability and the MTI in rendering the plasma unstable or stable. We derive exact stability criterion by working with the sharp boundary case in which the physical variables---temperature, mean molecular weight, density, and magnetic field---change discontinuously from one constant value to another on crossing the boundary.  Finally, we perform the linear stability analysis for the case of the differentially rotating plasma that is thermally and compositionally stratified as well. By assuming axisymmetric perturbations, we find the corresponding dispersion relation and the explicit mathematical expression determining the onset of the modified MTI.

\end{abstract}


\section{Introduction} \label{sec:intro}
\label{sec:Introduction}
In the presence of a dilute plasma
and a weak magnetic field in  galaxy clusters, the thermal conduction as well as the diffusion of ions are known to be strongly anisotropic
and occurring preferentially along the direction of the magnetic field.
While the magnetic field does not contribute significantly to the force balance in the equilibrium state,
it can lead to convective instabilities that arise from the perturbations about the background state in equilibrium. In such a weakly magnetized plasma, the stability
of a non-rotating plasma is dictated by the temperature gradient, whereas in a unmagnetized plasma, the stability depends on the entropy gradient.
With a temperature gradient parallel to the pressure gradient set by gravity and perpendicular to the background magnetic field, the convective instability that sets in is termed the magnetothermal instability (MTI)~\citep{balbus_apj00}. Analogously, the heat-flux-buoyancy-driven instability (HBI)~\citep{quataert_apj08} is said to be in action when the temperature gradient is parallel to the background magnetic field.

The plasma---intracluster medium (ICM)---pervading the galaxy cluster consists mostly of the hydrogen ions, the helium ions and the free electrons along with a small fraction 
of heavier elements. Studies have shown that the process of diffusion can lead to helium
sedimentation in the cluster core over a Hubble time, because of which, the composition can vary with the distance from the cluster center.  X-ray data gathered by the space observatories, the Chandra and the XMM-Newton, show that the temperature in the ICM decreases with radius 
outside the core~\citep{vikhlinin_apjl06, leccardi2008radial}. 
Although relatively under-explored, the mean molecular mass profile has been argued to decrease as well in the outer regions of the galaxy cluster~\citep{bulbul_aanda11}. Their study applies the theoretical model of~\citet{peng_apj09} for the helium sedimentation to the Chandra profile for the 
relaxed X-ray clusters. The timescale~\citep{fabian_mnras77, gilfanov_sal84, qin_apjl00,chuzhoy_mnras03} associated with the helium sedimentation in a typical cluster is of the order of  $10^{16}\, {\rm sec}$ which is about $40$-$50$ times smaller than the age of the universe.

This concentration gradient can feed the convective instabilities and thus has an effect on both the MTI and the HBI~\citep{pessah_apj13}. 
Further studies on this interplay between the temperature gradient and the concentration gradient has been well explored both numerically and analytically~\citep{berlok_apj15,berlok_apj16,berlok_apj16b, sadhu17mnras}. In a parallel investigation, 
researchers have drawn an analogy between  the convection in the ICM to the paradigmatic Rayleigh--B\'enard convection~\citep{himanshu_pla16}
and have found the critical value of the temperature gradient at which the MTI and the HBI are triggered. The analogy has been further extended to study the effect of the composition gradient on the critical value of temperature gradient required for the onset of the HBI~\citep{sadhu17mnras}. However, similar study of the MTI is still unreported. Additionally, it is also important to highlight that all the aforementioned investigations on the composition gradient induced instability in the ICM, assume the equilibrium state of the
ICM as the one in which the fluid velocity is zero. This is a drastic simplification because we know that almost all astrophysical bodies possess some amount of rotational motion. The rotational motion is most generally and realistically characterized by a differential rotation profile which means that the fluid is in a sheared state.

\citet{balbus_apj01} is probably the first study to consider the instability caused by the combined effect
of the MTI and the magneto-rotational instability (MRI). A study by~\citet{ren2010thermal} has looked at the combined effect of the MTI and the MRI
in a dissipative plasma and found that the instability condition with anisotropic viscosity
and resistivity is identical to that for a plasma without dissipative effects. They have obtained the same condition
as is in~\citet{balbus_apj01}. \citet{ren_pop11} have studied the effect of transverse shear, that could be as a result of differential rotation, on the MTI. They find that in the presence
of a transverse shear, a plasma which can resist the MTI may relent to the combined instability.

Many studies have found evidence for rotation in galaxy clusters.
Cosmological hydrodynmaical simulations indicate rotational pressure support in galaxy
clusters~\citep{lau2012constraining,fang2009rotation}.  Various observational measurements indicate the presence of rotation in the galaxy clusters. 
A large scale rotation has been found to be consistent with the observed ellipticity
in the X-ray isophotes~\citep{fang2009rotation}.
Most  galaxy clusters are ellipsoidal ~\citep{kalinkov05mnras,cooray00mnras} in shape,
complicating studies of resulting ellipticity in the isophotes.
The signature of  differential rotation 
in SC~0316-44 has been reported by~\citet{materne83aap} using a velocity gradient study.
\citet{oegerle92aj} have confirmed the presence of peculiar cD~galaxy velocity---
indicative of the rotation---in Abell 2017. For this same cluster,
\citet{kalinkov05mnras} have used a disk-like
model with  solid body rotation to understand the effects of global rotation,
and corrections to the virial mass. They  estimated the angular velocity
of  rotation of  Abell~2017 to be $\sim718\,{\rm km}\, {\rm s}^{-1} {\rm Mpc}^{-1}.$ 
The method of perspective rotation has been applied to the trace transverse motions of
 the Virgo cluster~\citep{hamden10apjl} and careful analyses suggest the presence of rotation in the cluster.
Apart from the optical and the positional data of galaxies, the rotation of  clusters can be verified with the help of the velocity distribution
and the dispersion of their members~\citep{tovmassian15ap}. The velocity structure and the large scale rotation of the ICM are however poorly constrained
from direct observations. Given the lack of constraints from observations, \citet{bianconi2013gas} have found, 
by assuming 
 cylindrical rotation profiles, that they could explain the observed ellipticity in the X-ray isophotes. 

In view of the presence and the importance of non-zero angular velocity, its effect on the dynamic stability of the ICM is worth studying.
To this end,~\citet{nipoti2013thermal},~\citet{nipoti_apj14}, and~\citet{nipoti15cup} have employed linear stability analysis to understand the instabilities possible in a rotating ICM in which magnetic field mediated anisotropic heat transport is active. However, the effect of composition gradient induced anisotropic diffusion on these instabilities remains hitherto unexplored. 

In what follows, first we introduce in section~\ref{sec:ge}, the governing equations for the fluid model of the ICM. Then, in section~\ref{sec:mti_cg}, we find out how the gradient of concentration affects the onset of the MTI in the presence of the magnetic tension. Subsequently, we investigate how the concentration gradient can counter the Kelvin--Helmholtz instability's (KHI) effect (section~\ref{sec:mti_kh}) and the differential rotation's effect (section~\ref{sec:mti_rot}) on the MTI. We end with section~\ref{sec:discussion} wherein we discuss the main conclusions and implications of the results obtained in this paper.
\section{Governing Equations and Relevant Definitions} \label{sec:ge}
\label{sec:RBA}
The continuum dynamics of the weakly-collisional dilute plasma having two species of ions (say, hydrogen and helium) is modelled by the following set of well-known coupled equations~(see also~\citet{pessah_apj13,himanshu_pla16,sadhu17mnras}: 
\begin{subequations}
\be &&\f{\pa \rho}{\pa t}+\bb{\nabla}\bcdot(\rho \mathbf{u})=0 \,,
\label{eq:rho}\\
&&\frac{\pa\mathbf{u}}{\pa t} + \mathbf{u} \bcdot \bb{\nabla}\mathbf{u} =-\f{1}{\rho}\bb{\nabla} \bcdot \left({\sf{P}} + 
\f{\bb{B}^2}{8\pi}{\sf{I}} - \f{{B^2}}{4\pi}\hat{\mathbf{b}}\hat{\mathbf{b}}\right) + \mathbf{g} \,, \,\,\,\,
\label{eq:v}\\
&&\f{\pa \mathbf{B}}{\pa t}=\bb{\nabla}\btimes(\mathbf{u}\btimes\mathbf{B})+ \eta \nabla^2\mathbf{B}\,,
\label{eq:b2}\\
&& \rho T\left(\frac{\pa s}{\pa t} + \mathbf{u} \bcdot \bb{\nabla}s \right)=
(p_\bot-p_\parallel)\f{d}{dt}\ln\f{B}{\rho^{\gamma-1}} 
+ \bb{\nabla} \bcdot \left[\mathbf{\chi\mathbf{\hat{b}}(\mathbf{\hat{b}}\bcdot
\bb{\nabla})T}\right],
\label{eq:S}\\ 
&&\frac{\pa c}{\pa t} + \mathbf{u} \bcdot \bb{\nabla}c = \bb{\nabla} \bcdot \left[\mathbf{D\mathbf{\hat{b}}(\mathbf{\hat{b}}\bcdot
\bb{\nabla})}c\right]
\,.
\label{eq:C}
\en
\end{subequations}
Here, $\bb{u}$, $\rho$, $T$, $c$, $\bb{B}$, $s$ respectively stand for the velocity, the density, the temperature,
the concentration of the helium ions, the magnetic field, and the specific entropy. 
Also, $\gamma$, $\eta$, $D$, and $\chi \approx 6 \times 10^{-7} T^{5/2} \,
\textrm{erg cm$^{-1}$ s$^{-1}$ K$^{-1}$}$~\citep{spitzer_book62,
	braginskii_rpp65} respectively denote the adiabatic index, the magnetic diffusivity, 
	the coefficient of the particle diffusion and the thermal conductivity.
The pressure tensor ${\sf{P}}=p_\bot {\sf{I}} + (p_\parallel - p_\bot) \hat{\mathbf{b}} 
\hat{\mathbf{b}}$~\citep{hollweg_jgr85}. Here  $\sf{I}$ is the identity matrix and, the subscripts
$\bot$ and $\parallel$ respectively refer to the directions perpendicular and parallel to the magnetic field.
We assume that the isotropic part of the pressure tensor  $P\equiv 2p_\bot/3 + p_\parallel/3$ 
satisfies the equation of state of an ideal gas, i.e., \be P=\frac{\rho k_B T}{\mu m_H},\en $k_B$, $\mu$, 
and $m_H$ being the Boltzmann constant, the mean molecular weight, and the atomic mass unit respectively. 
The concentration $c$ is related to the mean molecular weight through the following relationship:
$
\frac{1}{\mu}=(1-c)\left(\frac{1+Z_1}{\mu_1}\right)+c\left(\frac{1+Z_2}{\mu_2}\right),
$
where $Z_i$ and $\mu_i$ ($i\in\{1,2\}$) are respectively the atomic number and 
the molecular weight of the $i$th ion (either hydrogen or helium). The plasma-beta parameter is defined as $\beta\equiv{P}/({B^2/8\pi})$, where $P$ is the isotropic plasma pressure. 

For later usage, we also define the coefficient of thermal diffusion, $\lambda\equiv \chi/\rho c_p$;
the heat capacity at constant pressure, $c_p\equiv T({\partial s}/{\partial T})_P$; Chandrashekhar number, 
$Q\equiv B^2d^2/4\pi \rho \nu \eta$; Prandtl number, $P_r\equiv\nu/\lambda$; magnetic Prandtl number,
$P_m\equiv\nu/\eta$; Schwarzschild number, $\Sigma\equiv g\alpha Td/c_p\Delta T -1$; the ratio 
$\Lambda\equiv D/\lambda$; the thermal Rayleigh number, $R\equiv\alpha(\Delta T)gd^3/\lambda\nu$; and the
diffusion Rayleigh number, $R'\equiv\alpha_c(\Delta c)gd^3/\lambda\nu$. Here $\nu$, $d$, $\alpha$, and 
$\alpha_c\equiv(\partial\ln\rho/\partial \mu)_{P,T}$ are the kinematic viscosity, the height of the system, 
the coefficient of thermal expansion, and the coefficient of expansion
due to a change in the mean molecular weight respectively.
It may be noted that under the local approximation, as adopted in this paper, it has been implicitly assumed 
that the coefficients $\chi$ and $D$ are constants---an assumption which has been relaxed in,
e.g.,~ \citet{latter_mnras12, kunz_apj12, berlok_apj16}.
\section{MTI and concentration gradient}\label{sec:mti_cg}
After the linear stability analyses of the composition gradient modified HBI and MTI have been done
by~\citet{pessah_apj13} using the standard WKB-type approximations, \citet{sadhu17mnras}---following 
\citet{himanshu_pla16}---have shown how the problem can also be studied by employing the
technique~\citep{chandrashekhar_book81} traditionally used to study the convection problem in
a Rayleigh--B\'enard set-up. The latter technique has some added advantages, e.g., 
it connects with the boundary conditions usually adopted in the numerical simulations, it
helps one to find critical Rayleigh number corresponding to the HBI and the MTI, it facilitates easier 
analytical estimation of the contribution from the magnetic tension, 
and it may help us to come up with low dimensional models for convection, like the Lorenz model~\citep{lorenz63jas}. 

As mentioned earlier, the effect of the concentration gradient on the MTI has not been investigated as a
problem mapped onto the Rayleigh--B\'enard convection. In this section, we intend to do this investigation.
However, the primary goal of this section is to find the exact functional form of the coupling between the gradients of the temperature and the mean molecular weight that shows up in the criterion for the onset of the modified MTI. Additionally, we want to find the exact mathematical manner in which the magnetic tension affects the onset of the instability.

To this end, we model the ICM as a 
weakly collisional plasma consisting of the hydrogen and the helium ions (and, of course, free electrons) 
confined between two infinitely extended plane-parallel plates---\`a la the parallel-plane Rayleigh-B\'enard setup. {Here, the initial equilibrium state we  
consider is one in which the plasma is at rest and in  hydrostatic balance}. The acceleration due to gravity $\bb{g}$ is assumed to be in the vertical downward direction, i.e., $\bb{g}=-g\hat{\mathbf z}$. Also, it is assumed that the top and the bottom plates are fixed at constant temperatures $T_{\rm top}$ and ${T_{\rm bottom}}$ respectively, and at constant concentrations $c_{\rm top}$ and $c_{\rm bottom}$ respectively. An external uniform magnetic field, $\bb{B}$, is supposed to be along the $\hat{\mathbf x}$ direction, i.e., orthogonal to the gravity. For later convenience, we introduce two symbols: $\Delta T\equiv T_{\textrm{bottom}}-T_{\textrm{top}}$  and $\Delta c\equiv c_{\textrm{bottom}}-c_{\textrm{top}}$. 

The system under study follows the dynamics defined by the set of equations~(\ref{eq:rho})-(\ref{eq:C}) and the boundary conditions to be discussed a little later. In the stability analysis that follows in this section, an infinitesimal perturbation of a physical quantity $f$ about its equilibrium value is denoted by $\delta f$ which is supposed to have the form: $\delta f(\bb{x}, t)={\delta f}(z)\exp(-i\omega t+i\bb{k} \bcdot \bb{x})$ where $\bb{k}= (k_x, k_y, 0)$, $\omega$ is the complex frequency, and ${\delta f}(z)$ is the amplitude of the perturbation. We employ the  Boussinesq approximation since the sound speed is much greater than the growth rate of the linear instabilities of interest~\citep{balbus_apj00, balbus_apj01, quataert_apj08, pessah_apj13, himanshu_pla16}. 
This means that the density fluctuation in the momentum equation is neglected unless it is multiplied by $g$,
and the relative fluctuation in pressure ($\delta P/P$) is neglected in the energy equation. 
Also, the continuity equation is written as $\bb{\nabla}\bcdot\mathbf{u}=0$ in this approximation.

Closely following the algebraic steps detailed in~\citet{himanshu_pla16} and \citet{sadhu17mnras}, the linearized equations for the infinitesimal perturbations --- 
 $\delta u_z$, $\delta w_z$, $\delta B_z$, $\delta j_z$, $\delta\theta$, and $\delta c$ denoting the $z$-component of the momentum, the $z$-component of the vorticity, the $z$-component of magnetic field, the $z$-component of current density, the temperature, and the concentration perturbations respectively --- come out to be: 
\begin{subequations}
\be
&&\left[-i\omega(\partial_z^2-k^2)-\f{3}{k^2}k_x^4\partial_z^2\right]\delta u_z=-Rk^2\delta \theta + R'k^2 \delta c, \label{eq:smsvz}\\\nonumber 
&&+ ik_xQ\f{P_r}{P_m}(\partial_z^2-k^2)\delta B_z+\f{3}{k^2}k_x^3k_y\partial_z\delta w_z\,,\\
&&\left[-i\omega-\f{3}{k^2}k_x^2k_y^2\right]\delta w_z=ik_xQ\f{P_r}{P_m}\delta j_z + \f{3}{k^2}k_x^3k_y\partial_z\delta u_z\,,\label{eq:smswz}\\
&&(\partial_z^2-k^2+iP_m\omega)\delta B_z= -ik_x\f{P_m}{P_r}\delta u_z\,,\label{eq:smsbz}\\
&&(\partial_z^2-k^2+iP_m\omega)\delta j_z= -ik_x\f{P_m}{P_r}\delta w_z\,,\label{eq:smsjz}\\
&&(-k_x^2+iP_r\omega)\delta\theta=\Sigma\delta u_z+ik_x\delta B_z\,,\label{eq:smst}\\
&&(-\Lambda k_x^2+iP_r\omega)\delta c = -\delta u_z+ik_x\Lambda\delta B_z\,.\label{eq:smsc}
\en
\end{subequations}
As a check, we note that equations~(\ref{eq:smsvz})-(\ref{eq:smsc}) boil down to the corresponding correct equations for the case dealt with in~\citet{himanshu_pla16} where equation~(\ref{eq:C}) is absent. However, it must be kept in mind that equations (\ref{eq:smsvz})-(\ref{eq:smsc}) are non-dimensionalized. If prime ($'$) denotes the non-dimensional coordinates and variables, we have actually used in the equations: $\bb{x}'\equiv\bb{x}/d$, $t'\equiv t\nu/d^2$, $\delta u'_z\equiv\delta u_zd/\lambda$, $\delta w_{z}'\equiv\delta w_{z} d^2/\lambda$, $\delta B'_z\equiv\delta B_z/B$, $\delta j_{z}'\equiv\delta j_{z} d/B$, $\delta \theta'\equiv \delta T/\Delta T$, and $\delta c'\equiv\delta c/\Delta c$. However, without any loss of generality and for the sake of convenience, we have dropped the primes in equations~(\ref{eq:smsvz})-(\ref{eq:smsc}) and henceforth.

Out of many possible boundary conditions, we adopt the following boundary conditions~\citep{himanshu_pla16} at the bounding plates: $\delta u_z=0$, $\partial_z^2\delta u_z=0$, $\partial_z \delta w_z = 0$, $\delta B_z=0$, $\partial_z j_z=0$, $\delta \theta = 0$, and $\delta c=0$. These conditions physically imply that the normal component of the velocity must be zero on the boundary surfaces, surfaces are stress-free, boundaries are perfectly conducting, the boundary surfaces are at a constant temperature, and the surfaces are at constant concentration as well. While these are the most convenient conditions from an analytical point of view, these are certainly not the most general one. Although, one may loosely argue that being interested in the bulk of the plasma in a rather qualitative way, these boundary conditions suffices our requirement; this would not be a fully satisfactory justification either for choosing such convenient boundary conditions. Nevertheless, since some of these boundary conditions are numerically implementable in a straightforward manner~\citep{berlok_apj16,berlok_apj16b}, it makes our analytical results found using such boundary conditions benchmarks for such numerics.

At the condition of marginal stability, i.e., at the onset of the instability, mathematically, $\omega=0$. Since the typical values of $Q$ and ${\rm P}_{\rm m}$ for the 
ICM are of the order of $10^{24}-10^{40}$ and $10^{23}-10^{29}$~\citep{carilli_araa02, peterson_pr06, himanshu_pla16}, respectively, it is sensible to work with $Q,{\rm P_{\rm m}}\rightarrow\infty$.  In this limit, we combine equations (\ref{eq:smsvz})-(\ref{eq:smsc}) to arrive at the following single differential equation for $\delta u_z$:
\be
Q^2k_x^6(\partial_z^2-k^2)\delta u_z+Q(R-R')\f{{\rm P_{\rm m}}}{{\rm P_{\rm r}}}k^2k^4_x\delta u_z=0\,.\label{eq:Wmc2}
\en
Subsequent successive differentiations w.r.t. $z$, keeping in mind the condition that at boundaries ($z=0,1$) , $\delta u_z(z)=0$ and $\partial_z^2\delta u_z(z)=0$, leads to the following conditions at the boundaries
\be
\partial_z^{2m}\delta u_z=0\,\quad \textrm{with}\,\quad m\in\{0,1,2,...\}\,.
\en
Thus, the $n$th possible mode should have the form
\be
\delta u_z=A\sin n\pi z\,, 
\en
where $A$ is a constant. Therefore, equation (\ref{eq:Wmc2}) implies:
\be
&&R-R'=\f{k_x^2}{k^2}(n^2\pi^2+k^2)Q\f{{\rm P_{\rm r}}}{{\rm P_{\rm m}}},\,\textrm{where, }k_x\ne0\,;\label{eq:RQrel}\\
\Rightarrow&&-\f{d}{dz}\ln(T/\mu)=\f{k_x^2}{k^2}(n^2\pi^2+k^2)\left(\f{1}{\beta}\right)\,.\label{eq:RQrelaux}
\en
The minimum value for the gradient $-d\ln (T/\mu)/dz$ for the 
modes with $k_x, k_y\ne0$ to become marginally stable, is obtained from 
\be
\left.-\f{d}{dz}\ln(T/\mu)\right|_{\rm critical}&=&\min\left[\f{k_x^2(n^2\pi^2+k_x^2+k_y^2)}{k_x^2+k_y^2}\right]\left(\f{1}{\beta}\right)\,.
\label{eq:MTI_criterion_beta}
\en
The minimum value on the right hand side of equation~(\ref{eq:MTI_criterion_beta}) is determined by the mode under consideration. It should be borne in mind that we are working with a fluid model of plasma and our analysis is local. This means the wavenumbers ($k$) for which our stability analysis is plausible should be such that the corresponding wavelengths are much smaller than the system size and much larger than the mean free path of the ions. Similarly, although mathematically, $n = 1$ may be termed as the lowest mode, it violates the local approximation. This again is not a very serious objection because any other $n$ merely alters the multiplicative minimum in the right-hand side of equation~(\ref{eq:MTI_criterion_beta}) while keeping the $\beta$ dependence intact.

Linear stability analysis of the fluid model of the plasma can not bring forth a quantitative value of the minimum. Nevertheless, what we have achieved is that we have found out how the gradient of the temperature, the compositional gradient, and the plasma-$\beta$ (representative of the magnetic tension) are functionally connected at the marginal state of the system. We have explicitly proven that the threshold for the gradient of the logarithm of the ratio of the temperature to the mean molecular weight for the modified MTI to set in is inversely proportional to the plasma-$\beta$. In the outskirts of the galaxy clusters where the MTI is more probable, the temperature as well as the mean-molecular weight gradients are typically negative and thus, both the gradients may counterbalance the effects of each other.  

The X-ray observations from Chandra and XMM-Newton show that the temperature in the ICM decreases with radius  outside the core~\citep{vikhlinin_apjl06, leccardi2008radial}. The mean molecular weight for a homogeneous ICM with primordial abundance is roughly $0.6$. Studies~\citep{fabian_mnras77,rephaeli_apj1978,abramopoulos_apj1981,qin_apjl00,chuzhoy_mnras03,chuzhoy_mnras04} have shown that the process of diffusion can lead to the helium sedimentation in the cluster core over a Hubble time. Due to this the composition can vary with distance from the cluster center. 
Because of the fact that the helium is in a completely ionised form, emission line spectroscopy cannot be used to study the radial distribution of the helium ions. \citet{bulbul_aanda11} have applied the theoretical model of \citet{peng_apj09} for the helium sedimentation profile to the Chandra profile for 
relaxed X-ray clusters. They have used this to predict an upper limit to the effect of the helium sedimentation process on various physical quantities such as the gas mass profile, the total mass profile, and the scaling relations.
They have found a decrease of $25\%$ in the total mean molecular weight from the core of the cluster to the boundary of the cluster (defined at $r_{500}$, where $r_{500}$ is the radius enclosing a mean density of 500 in units of the critical density).
The temperature profile has been studied extensively in observations compared to the relatively unknown mean molecular mass profiles. The temperature profiles in observations show an average decrease of $\sim50\%$ from the core to $r_{500}$~\citep{arnaud2010universal}. These values indicate, at least for some typical ICM, the instability due to the compositional gradient may be comparable to that due to the unmodified MTI (see also Fig.~5 of  \citet{pessah_apj13}).

Before we end this section, let us consider modes with $k_x=0$, {without necessarily}
imposing the limits $Q,{\rm P_{\rm m}}\rightarrow\infty$. 
In this case, using equations (\ref{eq:smswz}), (\ref{eq:smst}), and (\ref{eq:smsc}), we arrive at
\be
-\rm{Pr}~\omega^2(\partial_z^2-k^2)\delta u_z=k_y^2(R\Sigma+R')\delta u_z.
\en
At the marginal state, $\omega=0$.
Using $\delta u_z=A\sin\pi z$ as the lowest mode, the condition for marginally stable state is $\left(R\Sigma+R'\right)=0$, i.e.,
\be
\f{g\alpha T d}{c_p}+\left(\f{\alpha_c}{\alpha }\right)\Delta c-\Delta T=0\,,
\en
and hence,
\be
&&\f{g\alpha d}{c_p}+\f{\Delta \mu}{\mu}-\f{\Delta T}{T}=0\,.
\label{eq:Schwarzschild}
\en
Equation (\ref{eq:Schwarzschild}) is the well known Ledoux criterion~\citep{ledoux47}.
\section{MTI, concentration gradient, and KHI}\label{sec:mti_kh}
\label{KH}

In the preceding section, $\bb{u}_0$---the value of the velocity of the plasma at equilibrium that we (henceforth) symbolize by the subscript `$0$'---is zero. But one of the main interests of this paper is to study the coupled effect of the instabilities fed by the temperature gradient, the concentration gradient, and the velocity shear. To this end, we use equations~(\ref{eq:rho})--(\ref{eq:C}) after ignoring the viscosity and the magnetic diffusivity, however magnetic field line mediated anisotropic conduction and diffusion are retained. Although this assumption is for the sake of analytical tractability, it may be noted that neither the viscosity nor the magnetic diffusivity appears in the criterion for the onset of the MTI (see equation~\ref{eq:RQrelaux}). This constraint allows us to work with only the isotropic part of the pressure tensor~\citep{hollweg_jgr85}. 

In this section, we work with a non-trivial equilibrium configuration of the ICM as described in this paragraph.
We take the fluid velocity to be directed along the $x$-axis with a shear perpendicular to it in the $z$-direction.
The magnetic field is parallel to the initial velocity with a shear in the $z$-direction as well. 
The thermodynamic variables in the equilibrium state only depend on the variable 
$z$ and hence the background gradients of the pressure, the density, the mean molecular weight, 
and the temperature are along the $z$-axis.
These gradients are thus parallel to each other, and perpendicular to the direction
of the mean velocity and the background magnetic field. Thus mathematically, the equilibrium state is given by:
\be
\rho_0=\rho_0(z),\, T_0=T_0(z),\,\mu_0=\mu_0(z),\, \bb{B_0}=B_0(z)\hat{x},\, \bb{u_0}=u_0(z)\hat{x}\,.
\en
 Additionally, the momentum equation gives the following constraint on the background equilibrium state: 
\be
&&\f{d}{dz}\left(P_0+\f{B^2_0}{8\pi} \right)=-\rho_0 g,
\en
where we have not ignored the magnetic pressure. We ensure, by construction, $\bb{B}_0\bcdot\bb{\nabla} T_0 = 0$, and this implies that the system is capable of sustaining the MTI besides the Kelvin--Helmholtz instability (KHI). Also, we set $\bb{B}_0\bcdot\bb{\nabla }\mu_0 = 0$~\citep{pessah_apj13}.  We recall that the KHI~\citep{helmholtz68pm,william71pm,chandrashekhar_book81} essentially refers to the finite velocity gradient induced instability witnessed in a freely ($g=0$) streaming shear flow. Of course, in case $g\ne0$ and the flow is stably stratified (just to avoid the Rayleigh--Taylor instability from setting in), shear flow can still facilitate the KHI if the velocity gradient is sharp enough. 

In view of the inhomogeneity in the $z$-direction, we assume infinitesimal perturbations of any of the physical variables, 
$f$ (say), of the form, $\delta f(x,y,z,t)=\delta f(z)\exp(-i\omega t + ik_xx+ik_yy)$ where $\omega$ is the frequency and ${\delta f}(z)$ is the amplitude of the perturbation. Since the mode of interest is the thermal convective mode, we impose~(\`a la~\citet{ren_pop11}): $\omega-k_xu_0\ll(\Gamma-1)(\bb{k}\bcdot\bb{b})^2\chi_cT_0/(\Gamma P_0)$. This inequality basically means that we are assuming that the perturbation evolves on a timescale much longer than the heat conduction timescale. The linearised versions of  the governing dynamical equations may be manipulated after lengthy but straightforward calculations to arrive at:
\be
\label{mode_eq}
&&\f{d}{dz}\left[\left(\rho_0(\omega-k_xu_0)^2-\f{k^2_xB^2_0}{4\pi} \right)\f{d}{dz}\bar{u}_z\right]\\\nonumber
&&\qquad=k^2\left[\left(\rho_0(\omega-k_xu_0)^2-\f{k^2_xB^2_0}{4\pi} \right)-g\rho_0\f{d}{dz}\ln\left(\f{T_0}{\mu_0}\right)\right]\bar{u}_z,\\\nonumber
\en
where $\bar{u}_z \equiv [\omega/({\omega-k_xu_0})]\delta u_z$. 
For a smooth shear flow normal mode analysis is known to fail and it is only through the concept of transient growth combined with nonlinear bootstrapping~\citep{trefethen39science} that one is able to reconcile with the fact that in reality smooth shear flow can actually become unstable. This is one of the reasons for adopting the sharp boundary model that we discuss now.
\subsection{Sharp boundary model}
It is not possible to solve equation~(\ref{mode_eq}) in its full generality. Consequently, in order to gain insight into the features of the instability that may result in the sheared ICM, we look for an analytically tractable case. The easiest one such case~\citep{ren_pop11} is probably in which the physical variables describing the system are considered to be constant on either side of $z=0$, where there is a discontinuous jump in the value of the variables. This jump is described by the use of the standard Heaviside function $h(z)$ as follows:
\be
\frac{T_0(z)-T_1}{T_2-T_1}=\frac{\mu_0(z)-\mu_1}{\mu_2-\mu_1}=\frac{\rho_0(z)-\rho_1}{\rho_2-\rho_1}=\frac{B_0(z)-B_1}{B_2-B_1}=h(z).
\en
Thus, integrating equation (\ref{mode_eq}) across the boundary yields
\be
\label{dispersgen}
\sum_{i=1}^2\rho_i(\omega - k_xu_i )^2 = gk \int_{0-}^{0+}\rho_0\frac{d}{dz}\ln(T_0/\mu_0)dz + k_x^2\frac{(B_1^2 + B_2^2)}{4\pi} \,.
\en 
We now consider two limiting cases as described below.
\subsubsection{Infinite-$\beta$ plasma}
The momentum equation integrated across the boundary gives, in the limit that $\beta \rightarrow \infty $, $\rho_1T_1/\mu_1 = \rho_2T_2/\mu_2$.  Hence, equation~(\ref{dispersgen}) becomes:
\be
\label{disperscase1}
\rho_2(\omega - k_xu_2 )^2 + \rho_1(\omega - k_x u_1 )^2 = -g k(\rho_2-\rho_1) + k_x^2\frac{(B_1^2 + B_2^2)}{4\pi}.
\en
This is a quadratic equation in $\omega$. The solution for $\omega$ has both real and imaginary parts.
The imaginary part that gives the growth rate is 
\begin{equation}
\gamma = \left[gkA_\rho + \frac{1}{4}k^2\cos^2\theta(1-A_\rho^2){\mid u_2-u_1 \mid}^2 -\bar{v}_A^2k^2\cos^2\theta\right]^{\frac{1}{2}}.
\end{equation} 
Here we have set 
$A_\rho \equiv (\rho_2 - \rho_1)/(\rho_1 + \rho_2)$, $\bar{v}_A \equiv \left[(B_1^2 + B_2^2)/4\pi(\rho_1 + \rho_2)\right]^{1/2}$, and $k_x = k\cos\theta$. Evidently, the system is unstable if, 
\begin{equation}
\frac{gA_\rho}{k \cos^2\theta} + \frac{1}{4}(1-A_\rho^2){\mid u_2-u_1 \mid}^2 > \bar{v}_A^2.
\end{equation}
This condition does not explicitly depend on the mean molecular weight. However, since $\rho_1T_1 /\mu_1=\rho_2T_2/\mu_2$,
we have $\rho_2 = \rho_1\mu_2T_1/\mu_1T_2$ making $A_\rho$ dependent
on $\mu_2/\mu_1$. If the temperature gradient and the concentration gradient
are of the same order, they tend to negate each other resulting in smaller values of $A_\rho$. It is easy to note that the critical value of $k$, at which the instability ensues, changes for the case where this is simultaneous presence of the concentration and the temperature gradients compared to the case where only the temperature gradient is in effect. 
\subsubsection{Finite $\beta$ and constant density}
If density is  constant and the assumption of infinite $\beta$ is relaxed, the value of $\gamma$ as obtained from equation~(\ref{disperscase1}) is given by
\begin{equation}
\gamma = \left[ \frac{1}{4}{\mid u_2-u_1 \mid}^2k_x^2\ + \frac{\;g\;k\;}{2}\;\ln\left(\frac{T_1/\mu_1}{T_2/\mu_2}\right)  -k_x^2\bar{v}_A^2\;\right]^{\frac{1}{2}}.
\end{equation}
Instability occurs if
\begin{equation}
 \frac{1}{4}{\mid u_2-u_1 \mid}^2 + \frac{g}{2k\cos^2\theta}\ln\left(\frac{T_1}{T_2}\right)  > \bar{v}_A^2+\frac{g}{2k\cos^2\theta}\ln\left(\frac{\mu_1}{\mu_2}\right)\,,\, (k_x\ne0).\label{inscond}
\end{equation}
In case there is no concentration gradient then this reduces to the condition found in~\citet{ren2013mri}. It is interesting to note that due to the positive definite contribution of the velocity shear (causing the KHI) in the left hand side of equation~(\ref{inscond}), an instability may arise even when the system is
otherwise stable to the MTI, i.e., when $T_1 < T_2$. 

The aforementioned coupling between the KHI and the MTI is further enriched by the effect of the composition gradient that may act as if to negate the effect of the temperature gradient. In case, the velocity shear is absent and the two gradients are of the same sign, then even when $T_1 > T_2$, the system may be stable provided  $({g}/{2k\cos^2\theta})\ln\left({T_1\mu_2}/{T_2\mu_1}\right)$ is not greater than the square of the modified Alfv\'{e}n velocity.
In the presence of the velocity shear, one may interpret equation~(\ref{inscond}) in another way: one may say that a negative concentration gradient ($\mu_1>\mu_2$) counters the destabilising effect of the KHI for a mode with wavenumber $k$.
\section{MTI, rotation, and the concentration gradient}\label{sec:mti_rot}
\label{contrib_rotat}
Now our intention is to theoretically study the effects of the rotation and the concentration gradients
on the MTI fed by the temperature gradient in the ICM. In this section, we find it convenient to work with a
cylindrical coordinate system---$(r,\phi,z)$. We assume a cylindrical rotation profile, i.e., $\Omega = \Omega(r)$. The Poincar\'{e}--Wavre theorem~\citep{tassoul1978theory} shows that a necessary and sufficient condition for $\Omega$ to be a function of $r$ only is that the isobaric and the isodensity surfaces coincide.
 These surfaces coincide
not only for barotropic fluids with $P = P(\rho)$ but also when $P = P(\rho, T, \mu)$ if the gradients of 
 temperature and density and $\mu$  are parallel to each other (pseudo-barotropic). 

We specifically consider the outer regions of the cluster where the MTI is potentially realizable.
In the outer regions of the cluster, radiative cooling is relatively unimportant and is not considered 
in this paper. In fact,~\citet{nipoti2013thermal} have concluded that in  astrophysically relevant configurations, 
radiative cooling is not crucial in fixing the stability criterion for the {weakly} magnetized plasma in rotation. Additionally, for the sake of simplicity (although not a particularly realistic model for the ICM), we set the azimuthal component of the magnetic field, $B_{0\phi}$, to zero. Since $\Omega = \Omega(r)$, the Ferraro isorotation law for steady axisymmetric flow---$\bb{B_0}\bcdot\bb{\nabla}\Omega = 0$~\citep{ferraro1937non}---implies that $B_{0r} = 0$. Together with
$\bb{B_0}\bcdot\bb{\nabla} T_0 = 0$, as is the traditionally required set up conducive to MTI, 
this implies $z$-independence of $T_0$. Consequently, the gradients of the thermodynamic variables
$T_0$,  $\mu_0$ , $\rho_0$, and $P_0$ are parallel to each other, and have only the radial components.

Subsequently, we linearize the equations of motion (\ref{eq:rho})--(\ref{eq:C}) assuming infinitesimal 
axisymmetric perturbations for every physical quantity $f$ (say) in the form, 
$\delta f(r,z)\propto\exp(-i\omega t + ik_rr+ik_zz)$ having a constant amplitude for the perturbation. 
We mention that, for the case in hand, we have ignored the viscosity and the magnetic diffusivity.
However the magnetic field line mediated anisotropic conduction and diffusion have been retained. 
The following two linearized equations, emanating from the entropy equation and the diffusion equation respectively, 
show the explicit effect of the concentration gradient while the other straightforward equations for the temporal evolution of the perturbations are not shown here as they are already known in the literature~\citep{nipoti2013thermal}:
 
\begin{subequations}	
	\begin{eqnarray}
	\frac{1}{\gamma}\left[i\gamma\omega\left(\frac{\rho}{\rho_0}\right) + u_r\left(\frac{\partial \ln P_0}{\partial r}-\gamma \frac{\partial \ln \rho_0}{\partial r}\right) + u_z\left(\frac{\partial \ln P_0}{\partial z}-\gamma \frac{\partial \ln \rho_0}{\partial z}\right)\right] \nonumber\\ 
	= -\frac{(\gamma-1)}{\gamma}\frac{\chi T}{P_0}(\bb{k}\bcdot\bb{b})^2 + i\left(\frac{\gamma-1}{\gamma}\right)\frac{\chi T_0}{P_0}\,\\\nonumber
	\times\left[(\bb{k}\bcdot\bb{b_0})\bb{\nabla} \ln T_0 + (\bb{b_0}\bcdot\bb{\nabla} \ln T_0)\bb{k}-2(\bb{k}\bcdot\bb{b_0})(\bb{b_0}\bcdot\bb{\nabla}\ln T_0)\bb{b_0}\right]\bcdot \bb{b}\,,\\\nonumber
	i\omega\left(\frac{\mu}{\mu_0}\right) - u_r\left(\frac{\partial \ln \mu_0}{\partial r}\right) - u_z\left(\frac{\partial \ln \mu_0}{\partial z}\right)= D(\bb{k}\bcdot\bb{b})^2\left(\frac{\mu}{\mu_0}\right) - iD\\
	\times\left[(\bb{k}\bcdot\bb{b_0})\bb{\nabla} \ln \mu_0 + (\bb{b_0}\bcdot\bb{\nabla}\ln\mu_0)\bb{k}-2(\bb{k}\bcdot\bb{b_0})(\bb{b_0}\bcdot\bb{\nabla}\ln\mu_0)\bb{b_0}\right]\bcdot\bb{b}\,.
	\end{eqnarray}
	\end{subequations}
The mathematical condition, that the linearized equations of the perturbations should have nontrivial solutions, yields the following dispersion relation: 
\be a_0\;\sigma^6 + a_1\;\sigma^5 + a_2\;\sigma^4 + a_3\;\sigma^3 + a_4\;\sigma^2 + a_5\;\sigma  + a_6 = 0\,,\label{eq:dispe}\en
where $\sigma\equiv -i\omega$ and the $a_i{\rm '}s$ are the real coefficients as defined below:
 \begin{subequations}
\begin{eqnarray} 
&&a_0 = 1\,,\\
&&a_1 = {\omega}_{c,a} + {\omega}_m\,, \\ 
&&a_2 = {\omega}_{BV}^2 + 2 {\omega}_A^2 + {\omega}_{c,a} {\omega}_m + {\omega}_{rot}^2\,,\\
&&a_3 = {\omega}_A^2 {\omega}_{c,mag} + 2 {\omega}_A^2 {\omega}_{c,a} + 2 {\omega}_A^2 {\omega}_m +{\omega}_{BV}^2 {\omega}_m + {\omega}_{c,a} {\omega}_{rot}^2\nonumber\\
&& + {\omega}_m {\omega}_{rot}^2 +{\omega}_{{\mu}p}^2 {\omega}_{c,a}\,, \\
&&a_4 =  -(4 k_z^2 \Omega^2 {\omega}_A^2)/k^2 + {\omega}_A^4 + {\omega}_A^2 {\omega}_{BV}^2  + {\omega}_A^2 {\omega}_{c,mag} {\omega}_m \nonumber\\ 
&&+ 2 {\omega}_A^2 {\omega}_{c,a} {\omega}_m +  {\omega}_A^2 {\omega}_{rot}^2 + {\omega}_{c,a} {\omega}_m {\omega}_{rot}^2 +  {\omega}_A^2 {\omega}_{{\mu},mag} {\omega}_{c,a}\,, \\ 
&&a_5 = {\omega}_A^4 {\omega}_{c,mag} + {\omega}_A^4 {\omega}_{c,a} + {\omega}_A^4 {\omega}_m  + {\omega}_A^2 {\omega}_{BV}^2 {\omega}_m \nonumber\\\nonumber
&&- (4 k_z^2 \Omega^2 {\omega}_A^2 ({\omega}_{c,a} + {\omega}_m))/k^2 + {\omega}_A^2 {\omega}_{c,a} {\omega}_{rot}^2 + {\omega}_A^2 {\omega}_m {\omega}_{rot}^2 \\
&&  +  {\omega}_A^2 {\omega}_{{\mu}p}^2 {\omega}_{c,a}\,,   \\
&&a_6 = {\omega}_A^4 {\omega}_{c,mag} {\omega}_m - (4 k_z^2 \Omega^2 {\omega}_A^2 {\omega}_{c,a}{\omega}_m)/k^2 + {\omega}_A^4 {\omega}_{c,a} {\omega}_m \nonumber\\
&&+ {\omega}_A^2 {\omega}_{c,a} {\omega}_m {\omega}_{rot}^2 + {\omega}_A^4 {\omega}_{{\mu},mag} {\omega}_{c,a}\,.\label{a6mid}
\end{eqnarray}
\end{subequations}
Here, with $\mathcal{D}$ defined as $\frac{k_r}{k_z}\frac{\partial}{\partial z} - \frac{\partial}{\partial r }$ and $K\equiv  \frac{4\pi P_0}{B_0^2}\frac{k_z^2}{k^2}\frac{\mathcal{D}\ln P_0}{k^2}$, the different frequencies' explicit expressions are as follow:
\begin{eqnarray}
&&\omega_{BV}^2 \equiv -\displaystyle\frac{k_z^2}{k^2}\frac{\mathcal{D} P_0 \mathcal{D} \ln P_0 \rho_0^{-\gamma} }{\rho_0 \gamma},\,\omega_{{\mu}p}^2\equiv \displaystyle\frac{k_z^2}{k^2}\frac{\mathcal{D} P_0 \mathcal{D} \mu_0}{\rho_0 \mu_0},
\omega_A^2 \equiv \displaystyle\frac{(\bf{k\cdot B})^2}{4\pi\rho_0},\qquad\nonumber\\
&&\omega_c \equiv \displaystyle\frac{(\gamma-1)}{\gamma}\displaystyle\frac{\chi T_0 k^2 }{P_0}, \,\omega_{rot}^2 \equiv -\displaystyle\frac{k_z^2}{k^2}\displaystyle\frac{\mathcal{D} (r^4\Omega^2)}{r^3},\,
\omega_{m} \equiv D \;(\bf{k\bcdot b})^2,\nonumber\\
&&\omega_{\mu,mag}\equiv -{Dk^2}K\left[\mathcal{D}\ln\mu_0 
		- 2(\bb{b_0}\bcdot\bb{\nabla}\ln\mu_0)\left(\displaystyle\frac{k_r}{k_z}b_{0z}-b_{or}\right)\right],\nonumber\\
&&\omega_{c,mag}\equiv {\omega_c}K\left[\mathcal{D}\ln T_0 
		- 2(\bb{b_0}\bcdot\bb{\nabla}\ln T_0)\left(\displaystyle\frac{k_r}{k_z}b_{0z}-b_{or}\right)\right],\nonumber\\
&&\omega_{c,a}\equiv\frac{\left(\bb{k}\bcdot\bb{b}\right)^2}{k^2}\omega_c.\nonumber
\end{eqnarray}		
We note that for the preceding polynomial equation~(\ref{eq:dispe}) in $\sigma$, the Routh--Hurwitz determinants~\citep{gantmacher.book} are given by,
\be
\Delta_j \equiv\det\begin{bmatrix} a_1 & a_3 & a_5 & a_7 & ... & a_{2j-1} \\ a_0 & a_2 & a_4 & a_6 & ... & a_{2j-2}\\0 &a_1 & a_3 & a_5  & ... & a_{2j-3}\\0 &a_0& a_2 & a_4  & ... & a_{2j-4}\\...&...&...&...&...&...\\0&0&0&0&...&a_j\end{bmatrix}
\en
where  $j = 1,\cdots,6$ and $a_i = 0 $ for $i > 6$. If none of the determinants are equal to zero, 
the Routh--Hurwitz theorem states that the 
system is stable if all the determinants are strictly positive. For the case in hand, straightforward but 
tedious calculations reveal that while $\Delta_1$ and $\Delta_2$ are always positive definite, $a_6>0$
is enough to ensure that $\Delta_3$, $\Delta_4$, $\Delta_5$, and $\Delta_6$ are simultaneously greater than zero. The explicit expressions of these determinants are given in Appendix~\ref{sec:AB}.

For the fluid considered here, $a_6$ [see equation~(\ref{a6mid})] is explicitly given by
\be
a_6 = \left[ 4 \pi \left\{\frac{d P_0}{dr}\left(-\frac{d\ln T_0}{dr} + \frac{d\ln \mu_0}{dr} \right)  + 2 \Omega r\frac{d\Omega}{dr} \rho_0\right\} \right.\nonumber \\ \left.+ B_0^2 k_z^2 \left(1+\frac{k_r^2}{k_z^2}\right)\right]\left[\frac{B_0^2 \chi D ( \gamma - 1) k_z^6 T_0 }{ 16 \pi^2 \gamma P_0 \rho_0^2 \left(1+\frac{k_r^2}{k_z^2}\right)}\right].
\en
Hence $a_6 > 0 $ implies 
\begin{equation}
 \frac{d P_0}{dr}\left(-\frac{d\ln T_0}{dr} + \frac{d\ln \mu_0}{dr} \right){+} \frac{d\Omega^2}{d\ln r} \,\rho_0 > -\frac{B_0^2 k_z^2}{4\pi} \left(1+\frac{k_r^2}{k_z^2}\right).
\end{equation}
Now, if stability has to be enforced for all values of $k_z$, then $a_6$ should be greater than zero for all values of $k_z$. Noting that $k_z$ cannot be zero in order to allow non-zero anisotropic heat conduction and diffusion, the condition
\begin{equation}
\label{mtcirot}
\frac{d\ln P_0}{dr}\left(-\frac{d\ln T_0}{dr} + \frac{d\ln \mu_0}{dr} \right) P_0 + \frac{d\Omega^2}{d\ln r}\;\rho_0\geq 0\,,
\end{equation}
necessarily and sufficiently ensures stability.

Thus, inequality (\ref{mtcirot}), in the absence of the concentration gradient, is the condition for stability when a temperature gradient and an angular velocity gradient are simultaneously
present \citep{balbus_apj01}. When the temperature gradient dominates, this implies that the gradients of the temperature and the pressure should be opposite to each other to ensure stability, otherwise the MTI ensues. In case the angular velocity gradient dominates, equation (\ref{mtcirot})
gives the condition for stability against the MRI~\citep{balbus1991powerful, balbus1992powerful}. In many of the astrophysical systems, the angular velocity decreases outwards and hence this stability condition is often violated. However, it is probably not
possible to make any universal quantitative comparison between the relative magnitudes of the two quantities.
One can only conclude that with the current knowledge of the
	velocity profiles in the ICM there is a non-negligible contribution from the MRI term (see Appendix~\ref{sec:AppA}) which can potentially modify the effects of the temperature gradient. The MRI term is sub-dominant in the outer regions of the cluster and may modify the dynamics near the temperature maximum at the edge of the core. It can be seen in appendix~\ref{sec:AppA} that there is a huge spread in the contribution of the MRI term depending on the three model configurations and hence we don't give a quantitative estimate regarding the relative magnitudes of the two quantities.
In its entirety, condition (\ref{mtcirot}) is the condition for stability when, in addition to the temperature gradient, the gradient in composition and a differential rotation are present simultaneously. 
Since the ion diffusion, just like diffusion of heat, preferentially occurs along the magnetic field lines,
the effect of the concentration gradient appears in equation~(\ref{mtcirot}) in a functionally similar fashion
to the temperature-term. This is something we have seen in the earlier sections of this paper as well. It however must be mentioned that this similarity is also the consequence of adopting ideal gas law for the plasma. As far as the importance of this term is concerned, it has been already indicated that the contribution of the concentration gradient can be comparable
to the contribution from the MTI term, and hence it can't be ignored. In the outer part of the galaxy cluster it can counteract the MRI term's effect in bringing about instability.

\begin{figure}
	\centering
	\includegraphics[scale=0.36]{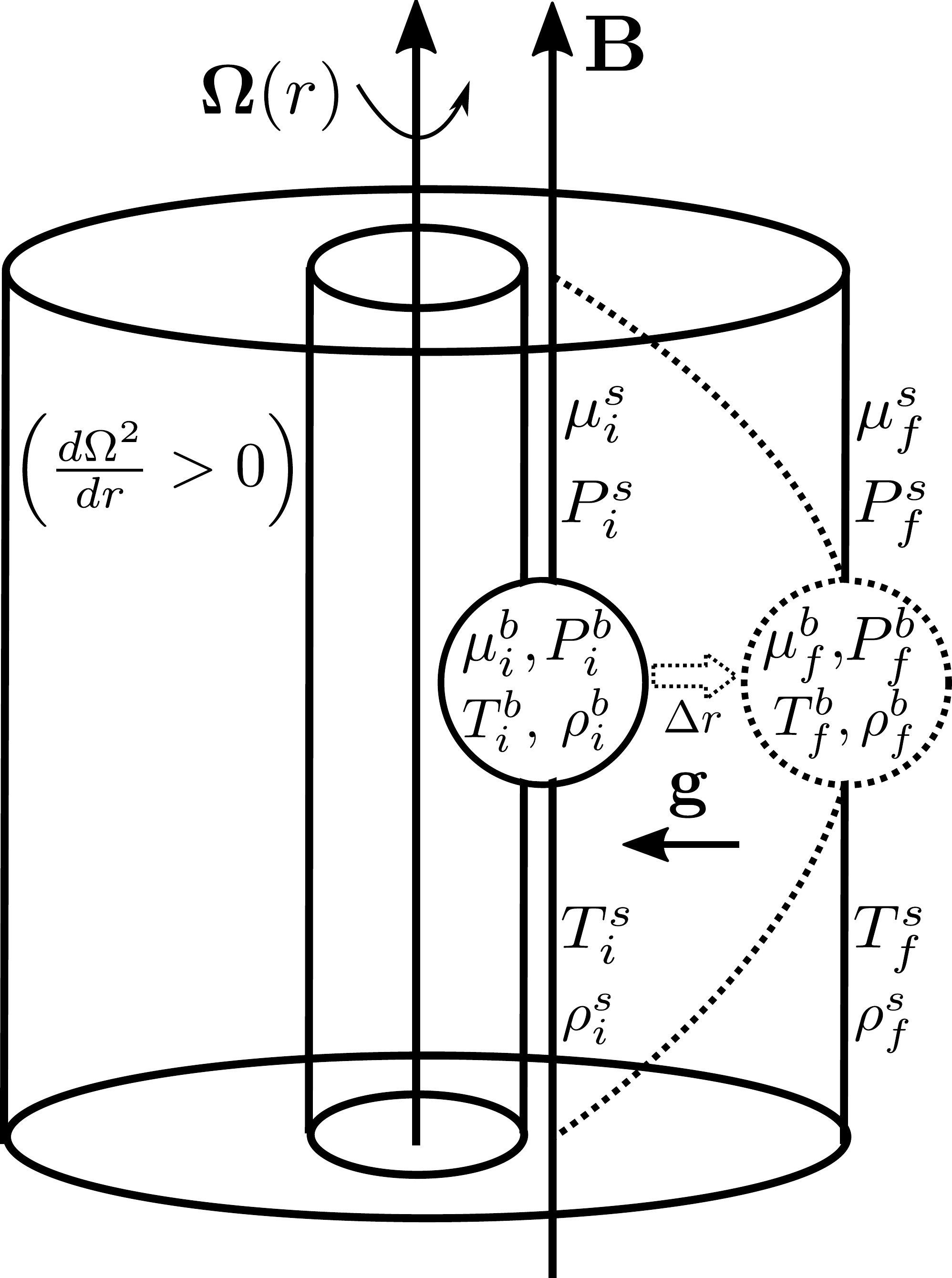}
	\caption{Virtual outward displacement of a fluid blob (circle with solid boundary) outwards to a new position (depicted by dashed circle) in a radially stratified, rotating ICM.}
	\label{fig:blob_mti}
\end{figure}
{Before we go to the concluding section, let us revisit stability criterion~(\ref{mtcirot}) but now in more physical and intuitive terms. Consider figure~\ref{fig:blob_mti} where we show a radially stratified plasma differentially rotating such that $d\Omega^2/dr>0$. Gravity has a radial component. If a fluid blob (or element or parcel) is displaced from the inner part of the plasma to the outer part, the positive differential rotation causes the blob to come back to its original position, i.e., fluid is stable against MRI; this has been already beautifully explained~\citep{balbus03araa} by considering two fluid elements connected by a spring-like force due to the presence of magnetic field lines. We, thus, focus solely on the concentration modified MTI part of inequality~(\ref{mtcirot}). One way to doing that would be to set $\Omega={\rm constant}$, giving $d(T/\mu)/dr>0$ as the criterion since $d P/dr\propto-g$ with positive proportionality factor (here and henceforth, we have suppressed subscript ``0" for notational convenience). In order to physically understand why stability is synonymous with $d(T/\mu)/dr>0$, we note that the (virtually) displaced fluid blob carries the magnetic field line---originally attached to the blob---along with it to the new position. We denote the pressure, the density, the temperature, and the mean molecular weight of the fluid parcel at initial position by $P_i^b$, $\rho_i^b$, $T_i^b$, and $\mu_i^b$, while that of the surrounding as $P_i^s$, $\rho_i^s$, $T_i^s$, and $\mu_i^s$. After a displacement $\Delta r$, the relevant thermodynamical variables of the blob and the surrounding become ($P_f^b$, $\rho_f^b$, $T_f^b$, $\mu_f^b$) and ($P_f^s$, $\rho_f^s$, $T_f^s$, $\mu_f^s$) respectively. The pressure difference between the blob and its immediate surrounding vanishes quickly owing to acoustic waves~\citep{chandran_apj06} whose timescale is much smaller than the timescales associated with the modes related to MTI. Hence the pressures inside and outside of the parcel become equal i.e., $P_f^b=P_f^s$. Additionally, the magnetic field line carried by the blob to the new position has its ends still rooted to the surrounding of the initial position. The magnetic field mediated conduction and diffusion, thus, forces $T_f^b=T_i^s$ and $\mu_f^b=\mu_i^s$.
Therefore, recalling the ideal gas equation, we get:
\be
&&\frac{\rho_f^b T_i^s}{\mu_i^s}=\frac{\rho_f^s T_f^s}{\mu_f^s}\,.\label{eq:idealgas}
\en
Since $dT/dr$ and $d\mu/dr$ are the gradients of temperature and mean-molecular weight respectively, equation (\ref{eq:idealgas}) becomes
\be
&&\frac{\rho_f^b T_i^s}{\mu_i^s}=\rho_f^s\frac{T_i^s+\frac{dT}{dr}\Delta r}{\mu_i^s+\frac{d\mu}{dr}\Delta r}\,.\label{eq:idealgrad}
\en
Keeping the terms up to linear in $\Delta r$ leads to 
\be
&&\rho_f^b-\rho_f^s \propto \frac{d}{dr}\ln(T/\mu)\,,
\en
having a positive proportionality factor.
Hence we conclude that the displaced blob sinks back to its original position if $d\ln(T/\mu)/dr\textgreater0$ because blob is denser than its immediate surrounding at the displaced position. This shows that stability of the system against infinitesimal perturbation is synonymous with inequality~(\ref{mtcirot}) when one sees the aforementioned physical picture in conjunction with the physical explanation of the MRI~\citep{balbus03araa}. 
\section{Discussions and Conclusions} \label{sec:discussion}
\label{sec:DaC}
Earlier investigations into the effect of the concentration gradient induced instabilities on the MTI~\citep{pessah_apj13,berlok_apj15,berlok_apj16} do not consider the non-trivial background velocity profile of the ICM. As elaborated in the introduction of this paper, the simulations and the observations support the view that ICM plasma, that is in a sheared state, is most generally and realistically characterized by a differential rotation. Additionally it has been argued that the gradients of the temperature and the mean molecular
 mass can be comparable in the outer regions of galaxy clusters.
 
These observations have motivated us, in this paper, to study the instability brought forth by the simultaneous presence of the MTI, the velocity shear, and the concentration gradient. In due course, we have also understood how the concentration gradient affects a weakly collisional plasma that is under the influence of the MRI and the MTI. Even though, for the sake of simplicity, we have confined our studies to the case of plasmas where the gradients of $P$, $\mu$, $\rho$, and $T$ are parallel to each other, 
we believe that our study very transparently clarifies the importance of the variation of the mean molecular weight---possibly due to the process of helium sedimentation---on the stability of the ICM.

While studying how the compositional gradient changes the linear stability criterion for the instability that is the combined instability consisting of the KHI and the MTI, we have focussed on the two special cases: the infinite $\beta$ case
and the case where density is constant. Although this has been done for the sake of the analytical tractability, we have provided the full equation (equation (\ref{mode_eq})) governing the linear evolution of the perturbation $\delta u_z$. In principle, one can analyse that equation numerically without simplifying it further. However, the two cases we have detailed are not only sufficient for our purpose but also insightful. We may recall that $\beta$ values in the outskirts of the galaxy cluster can be quite high ($\gtrsim10^4$) and hence the aforementioned infinite $\beta$ case is very apt. We have found that, as expected, the gradients of the mean molecular weight and the temperature couple together in deciding the instability criterion. We have also found that the critical value of $k$, at which the onset of instability occurs, changes.

The ion diffusion, just like the diffusion of heat, takes place in an anisotropic way in the ICM (obeying ideal gas law)---preferentially along the direction of the magnetic field. Hence there is a similarity in the manner in which the temperature gradient and the concentration gradient appear
in the conditions for stability in all the cases studied herein. Specifically, we note that wherever in mathematical equations, $\ln T$ appears in the absence of the concentration gradient, in the presence of concentration gradient induced diffusion $\ln T$ always seem to appear in conjunction with a $-\ln\mu$ term.

It must be remarked that while incorporating the effects of rotation on the MTI modified by the presence of a compositional gradient, we have specifically worked with axisymmetric perturbations. Of course, calculations with any non-axisymmetric perturbation are much more involved and can change the stability criterion. However, it is curious to note that if the magnetic field is set to zero then although the axisymmetric perturbations may or may not be stabilized by the differential rotation, the non-axisymmetric perturbations are susceptible to be stabilized by the differential rotation~\citep{nipoti2013thermal}. One may also note that we have not included radiative cooling in our calculations. Its inclusion is, however, not important for what we want to achieve in this paper: we want to analytically establish that the diffusion of the helium ions due to the presence of the compositional gradient can counter the influence of the temperature gradient and the velocity shear (possibly caused by differential rotation) on the MTI.  Moreover, after detailed discussions \citet{nipoti2013thermal} conclude that {``...radiative cooling is not the key factor determining the instability of a rotating magnetized plasma.''}

We conclude by commenting that with this paper, we have completed the linear stability analyses of  the fluid model of the weakly collisional dilute plasma, say ICM, that is a binary mixture of the hydrogen and the helium ions (along with the free electrons). This scheme of research had started with~\citet{pessah_apj13} and continued in a few subsequent papers~\citep{berlok_apj15,berlok_apj16,berlok_apj16b,sadhu17mnras}. 
%

\section*{Acknowledgements} S.C. thanks Martin E. Pessah for helpful discussions and gratefully acknowledges financial support from the INSPIRE faculty fellowship (DST/INSPIRE/04/2013/000365) awarded by the Department of Science and Technology, India.

\bibliographystyle{mn2e}
\bibliography{Gupta_etal_bibliography}

\begin{thebibliography}{56}
\expandafter\ifx\csname natexlab\endcsname\relax\def\natexlab#1{#1}\fi

\bibitem[{{Abramopoulos}, {Chanan} \& {Ku}(1981){Abramopoulos}, {Chanan}, \&
  {Ku}}]{abramopoulos_apj1981}
{Abramopoulos} F., {Chanan} G.~A., {Ku} W.~H.-M., 1981, ApJ, 248, 429

\bibitem[{Arnaud {et~al}\mbox{.}(2010)Arnaud, Pratt, Piffaretti, B{\"o}hringer,
  Croston, \& Pointecouteau}]{arnaud2010universal}
Arnaud M., Pratt G., Piffaretti R., B{\"o}hringer H., Croston J., Pointecouteau
  E., 2010, Astronomy \& Astrophysics, 517, A92

\bibitem[{Balbus(2000)}]{balbus_apj00}
Balbus S.~A., 2000, ApJ, 534, 420

\bibitem[{Balbus(2001)}]{balbus_apj01}
Balbus S.~A., 2001, ApJ, 562, 909

\bibitem[{Balbus(2003)}]{balbus03araa}
Balbus S.~A., 2003, Annual Review of Astronomy and Astrophysics, 41, 555

\bibitem[{Balbus \& Hawley(1991)}]{balbus1991powerful}
Balbus S.~A., Hawley J.~F., 1991, The Astrophysical Journal, 376, 214

\bibitem[{Balbus \& Hawley(1992)}]{balbus1992powerful}
Balbus S.~A., Hawley J.~F., 1992, The Astrophysical Journal, 400, 610

\bibitem[{Berlok \& Pessah(2015)}]{berlok_apj15}
Berlok T., Pessah M.~E., 2015, ApJ, 813, 22

\bibitem[{Berlok \& Pessah(2016{\natexlab{a}})}]{berlok_apj16}
Berlok T., Pessah M.~E., 2016{\natexlab{a}}, ApJ, 824, 32

\bibitem[{Berlok \& Pessah(2016{\natexlab{b}})}]{berlok_apj16b}
Berlok T., Pessah M.~E., 2016{\natexlab{b}}, ApJ, 833, 164

\bibitem[{Bianconi, Ettori \& Nipoti(2013)Bianconi, Ettori, \&
  Nipoti}]{bianconi2013gas}
Bianconi M., Ettori S., Nipoti C., 2013, Monthly Notices of the Royal
  Astronomical Society, 434, 1565

\bibitem[{Braginskii(1965)}]{braginskii_rpp65}
Braginskii S.~I., 1965, RvPP, 1, 205

\bibitem[{Bulbul {et~al}\mbox{.}(2011)Bulbul, Hasler, Bonamente, Joy, Marrone,
  Miller, \& Mroczkowski}]{bulbul_aanda11}
Bulbul G.~E., Hasler N., Bonamente M., Joy M., Marrone D., Miller A.,
  Mroczkowski T., 2011, A\&A, 533, A6

\bibitem[{{Carilli} \& {Taylor}(2002)}]{carilli_araa02}
{Carilli} C.~L., {Taylor} G.~B., 2002, ARA\&A, 40, 319

\bibitem[{{Chandran} \& {Dennis}(2006)}]{chandran_apj06}
{Chandran} B.~D., {Dennis} T.~J., 2006, ApJ, 642, 140

\bibitem[{{Chandrashekhar}(1981)}]{chandrashekhar_book81}
{Chandrashekhar} S., 1981, Hydrodynamic and Hydromagnetic Stability. Dover
  Publications, NY

\bibitem[{{Chuzhoy} \& {Loeb}(2004)}]{chuzhoy_mnras04}
{Chuzhoy} L., {Loeb} A., 2004, MNRAS, 349, L13

\bibitem[{{Chuzhoy} \& {Nusser}(2003)}]{chuzhoy_mnras03}
{Chuzhoy} L., {Nusser} A., 2003, MNRAS, 342, L5

\bibitem[{{Cooray}(2000)}]{cooray00mnras}
{Cooray} A.~R., 2000, \mnras, 313, 783

\bibitem[{{Fabian} \& {Pringle}(1977)}]{fabian_mnras77}
{Fabian} A.~C., {Pringle} J.~E., 1977, MNRAS, 181, 5P

\bibitem[{Fang, Humphrey \& Buote(2009)Fang, Humphrey, \&
  Buote}]{fang2009rotation}
Fang T., Humphrey P., Buote D., 2009, The Astrophysical Journal, 691, 1648

\bibitem[{Ferraro(1937)}]{ferraro1937non}
Ferraro V.~C., 1937, Monthly Notices of the Royal Astronomical Society, 97, 458

\bibitem[{Gantmacher(1959)}]{gantmacher.book}
Gantmacher F., 1959, Applications of the Theory of Matrices. Interscience
  Publishers, New York

\bibitem[{{Gil'fanov} \& {Syunyaev}(1984)}]{gilfanov_sal84}
{Gil'fanov} M.~R., {Syunyaev} R.~A., 1984, Soviet Astronomy Letters, 10, 137

\bibitem[{Gupta {et~al}\mbox{.}(2016)Gupta, Rathor, Pessah, \&
  Chakraborty}]{himanshu_pla16}
Gupta H., Rathor S., Pessah M., Chakraborty S., 2016, PhLA, 380, 2407

\bibitem[{Hamden {et~al}\mbox{.}(2010)Hamden, Simpson, Johnston, \&
  Lee}]{hamden10apjl}
Hamden E.~T., Simpson C.~M., Johnston K.~V., Lee D.~M., 2010, The Astrophysical
  Journal Letters, 716, L205

\bibitem[{Helmholtz(1868)}]{helmholtz68pm}
Helmholtz H., 1868, Philosophical Magazine, 36, 337

\bibitem[{Hollweg(1985)}]{hollweg_jgr85}
Hollweg J.~V., 1985, JGRA, 90, 7620

\bibitem[{Kalinkov {et~al}\mbox{.}(2005)Kalinkov, Valchanov, Valtchanov,
  Kuneva, \& Dissanska}]{kalinkov05mnras}
Kalinkov M., Valchanov T., Valtchanov I., Kuneva I., Dissanska M., 2005,
  Monthly Notices of the Royal Astronomical Society, 359, 1491

\bibitem[{{Kunz} {et~al}\mbox{.}(2012){Kunz}, {Bogdanovi{\'c}}, {Reynolds}, \&
  {Stone}}]{kunz_apj12}
{Kunz} M.~W., {Bogdanovi{\'c}} T., {Reynolds} C.~S., {Stone} J.~M., 2012, ApJ,
  754, 122

\bibitem[{{Latter} \& {Kunz}(2012)}]{latter_mnras12}
{Latter} H.~N., {Kunz} M.~W., 2012, MNRAS, 423, 1964

\bibitem[{Lau {et~al}\mbox{.}(2012)Lau, Nagai, Kravtsov, Vikhlinin, \&
  Zentner}]{lau2012constraining}
Lau E.~T., Nagai D., Kravtsov A.~V., Vikhlinin A., Zentner A.~R., 2012, The
  Astrophysical Journal, 755, 116

\bibitem[{Leccardi \& Molendi(2008)}]{leccardi2008radial}
Leccardi A., Molendi S., 2008, Astronomy \& Astrophysics, 486, 359

\bibitem[{{Ledoux}(1947)}]{ledoux47}
{Ledoux} P., 1947, ApJ, 105, 305

\bibitem[{Lorenz(1963)}]{lorenz63jas}
Lorenz E.~N., 1963, J. Atmospheric Sciences, 20, 130

\bibitem[{{Materne} \& {Hopp}(1983)}]{materne83aap}
{Materne} J., {Hopp} U., 1983, \aap, 124, L13

\bibitem[{Nipoti \& Posti(2013)}]{nipoti2013thermal}
Nipoti C., Posti L., 2013, Monthly Notices of the Royal Astronomical Society,
  428, 815

\bibitem[{{Nipoti} \& {Posti}(2014)}]{nipoti_apj14}
{Nipoti} C., {Posti} L., 2014, ApJ, 792, 21

\bibitem[{Nipoti {et~al}\mbox{.}(2015)Nipoti, Posti, Ettori, \&
  Bianconi}]{nipoti15cup}
Nipoti C., Posti L., Ettori S., Bianconi M., 2015, Journal of Plasma Physics,
  81

\bibitem[{{Oegerle} \& {Hill}(1992)}]{oegerle92aj}
{Oegerle} W.~R., {Hill} J.~M., 1992, \aj, 104, 2078

\bibitem[{{Peng} \& {Nagai}(2009)}]{peng_apj09}
{Peng} F., {Nagai} D., 2009, ApJ, 693, 839

\bibitem[{Pessah \& Chakraborty(2013)}]{pessah_apj13}
Pessah M.~E., Chakraborty S., 2013, ApJ, 764, 13

\bibitem[{Peterson \& Fabian(2006)}]{peterson_pr06}
Peterson J., Fabian A., 2006, PhR, 427, 1

\bibitem[{Qin \& Wu(2000)}]{qin_apjl00}
Qin B., Wu X.-P., 2000, ApJL, 529, L1

\bibitem[{Quataert(2008)}]{quataert_apj08}
Quataert E., 2008, ApJ, 673, 758

\bibitem[{{Ren} {et~al}\mbox{.}(2011){Ren}, {Cao}, {Dong}, {Wu}, \&
  {Chu}}]{ren_pop11}
{Ren} H., {Cao} J., {Dong} C., {Wu} Z., {Chu} P.~K., 2011, PhPl, 18, 022110

\bibitem[{Ren {et~al}\mbox{.}(2013)Ren, Cao, Li, \& Chu}]{ren2013mri}
Ren H., Cao J., Li D., Chu P.~K., 2013, Physics of Plasmas, 20, 032102

\bibitem[{Ren {et~al}\mbox{.}(2010)Ren, Wu, Dong, \& Chu}]{ren2010thermal}
Ren H., Wu Z., Dong C., Chu P.~K., 2010, Physics of Plasmas, 17, 052102

\bibitem[{Rephaeli(1978)}]{rephaeli_apj1978}
Rephaeli Y., 1978, ApJ, 225, 335

\bibitem[{Sadhukhan, Gupta \& Chakraborty(2017)Sadhukhan, Gupta, \&
  Chakraborty}]{sadhu17mnras}
Sadhukhan S., Gupta H., Chakraborty S., 2017, Monthly Notices of the Royal
  Astronomical Society, 469, 2595

\bibitem[{Spitzer(1962)}]{spitzer_book62}
Spitzer L., 1962, Physics of Fully Ionized Gases. Wiley Interscience, New York

\bibitem[{{Tassoul}(1978)}]{tassoul1978theory}
{Tassoul} J.-L., 1978

\bibitem[{Thomson(1871)}]{william71pm}
Thomson W., 1871, Philosophical Magazine, 42, 362

\bibitem[{Tovmassian(2015)}]{tovmassian15ap}
Tovmassian H.~M., 2015, Astrophysics, 58, 471

\bibitem[{Trefethen {et~al}\mbox{.}(1993)Trefethen, Trefethen, Reddy, \&
  Driscoll}]{trefethen39science}
Trefethen L.~N., Trefethen A.~E., Reddy S.~C., Driscoll T.~A., 1993, Science,
  261, 578

\bibitem[{Vikhlinin {et~al}\mbox{.}(2006)Vikhlinin, Kravtsov, Forman, Jones,
  Markevitch, Murray, \& Speybroeck}]{vikhlinin_apjl06}
Vikhlinin A., Kravtsov A., Forman W., Jones C., Markevitch M., Murray S.~S.,
  Speybroeck L.~V., 2006, ApJ, 640, 691

\end{thebibliography}
\appendix
\section{Numerical Estimates: Relative importance of the MRI term}
\label{sec:AppA}
In order to get a quantitative idea of how the two terms (see equation~\ref{mtcirot})---one leading to the MTI and the other to the MRI---compare in the context of the galaxy clusters, we find estimates based on the results present in the current research literature. The angular velocity profiles used by~\citet{bianconi2013gas} could explain the observed ellipticity in the X-ray isophotes.
These are derived from two separate velocity profiles with free parameters for a scale radius and a velocity scale.
These velocity profiles are given by:
\begin{equation}
\label{vmodel1}
u_{0\phi}^2 = u_\star^2\;\left[\frac{\ln(S+1)}{S} - \dfrac{1}{S + 1}\right],
\end{equation}
and
\begin{equation}
\label{vmodel2}
u_{0\phi}^2 = u_\star^2\;\dfrac{S^2}{(S+1)^4},
\end{equation}
where $S = {R}/{R_0}$, $R_0$ being a scale radius. They found that equation~(\ref{vmodel1}) with $u_\star = 1120\, {\rm km} \,{\rm s}^{-1}$ and $R_0 = 170 \,{\rm kpc}$;
and equation~({\ref{vmodel2}}) with $u_\star = 2345 \, {\rm km} \,{\rm s}^{-1}$ and $R_0= 120 \,{\rm kpc}$
as well as with  $u_\star= 1000 \, {\rm km} \,{\rm s}^{-1}$ and $R_0 = 120 \,{\rm kpc}$ produced ICM ellipticity
comparable to that from observations.
The values of the slope ${d\Omega^2}/{dr}$  for these three velocity profiles at $r_{500}$  are respectively 
(in units of $10^{-3} \,{\rm km}^2\, {\rm s}^{-2}\,{\rm kpc}^{-3}$) $-0.20$, $ -0.05$, and  $-0.01$.  
In comparison, the values at $0.2\,r_{500}$ ($\sim$radius of the core) are $-25.75$,  $-35.56$, and $-6.47$
respectively in the same unit. For these values and the mean observed temperature and pressure profiles for galaxy clusters,
an estimate of the relative contribution from the temperature gradient term and
differential rotation term in equation (\ref{mtcirot}) is now made,
i.e., we estimate using these values the relative contributions of the terms:
$\frac{d\ln P_0}{dr}\left(-\frac{d\ln T_0}{dr} \right) P_0 $ and  $\frac{d\Omega^2}{d\ln\,r}\;\rho_0 $.
Dividing both terms by $P_0$, these two terms are:
$\frac{d\ln P_0}{dr}\left(-\frac{d\ln T_0}{dr} \right) $ and  $\frac{d\Omega^2}{d\ln\,r}\;\frac{\rho_0}{P_0} $ respectively.
\citet{arnaud2010universal} gives a universal
pressure profile for galaxy clusters. Using the mean temperature profile and the pressure profile
as found from observations (see Figs. 2 and 3 in~\citet{arnaud2010universal} respectively),
the slopes for the pressure and temperature profile at $0.2\,r_{500}$ and $r_{500}$ are estimated. These values for the slopes together
with the values for ${d\Omega^2}/{dr}$ mentioned above as well as a typical cluster temperature (proportional to
${\rho_0}/{P_0}$) give the required estimates for the two terms for the three model estimates for ${d\Omega^2}/{dr}$.
At $r_{500}$, the estimates are $-2.5\times10^{-6}\; {\rm kpc}^{-2}$ for the first term; and $-2.1\times10^{-7}\; {\rm kpc}^{-2}$, $-4.9\times10^{-8}\; {\rm kpc}^{-2}$, and 
$-8.9\times10^{-9}\; {\rm kpc}^{-2}$ for the second term for three values of ${d\Omega^2}/{dr}$ respectively.
At $0.2\,r_{500}$, the similar values are $-2.1\times10^{-7}\; {\rm kpc}^{-2}$ for the first term; and $-5.4\times10^{-6}\; {\rm kpc}^{-2}$, $-7.4\times10^{-6}\; {\rm kpc}^{-2}$, and $
-1.4\times10^{-6}\; {\rm kpc}^{-2}$~respectively. Thus, even within the three model-estimates for ${d\Omega^2}/{dr}$, one can notice a substantial spread of at least two orders of magnitude and, thus, it is not
possible to make any universal quantitative estimate regarding the relative magnitudes of the two quantities. 
\section{The Routh--Hurwitz determinants}
\label{sec:AB}
Here we provide the explicit expressions of the Routh--Hurwitz determinants as used in this paper: 
\begin{eqnarray}
&&\Delta_1 = \frac{\left(\bb{k}\bcdot\bb{b}\right)^2(\gamma-1)\chi T_0 }{\gamma P_0} + D(\mathbf{k}\cdot\mathbf{b})^2\,,\\
&&\Delta_2=
\frac{ \chi \,( \gamma-1 )\, k_z^2\, T_0 }{\gamma^2\, P_0^2 \,\rho_0 \,\left(1+\frac{k_r^2}{k_z^2}\right)}\bigg[(\gamma-1)\,\left(\frac{\partial P_0}{\partial r}\right)^2 \nonumber\\
&&\,+\,D\,\rho_0\;k_z^4\;\left(D\;\gamma P_0\;+\chi\,(\gamma-1)\,T_0\,\right)\left(1+\frac{k_r^2}{k_z^2}\right)\bigg]\,,
\end{eqnarray}
\begin{eqnarray}
&&\Delta_3=
 \frac{1}{(\gamma^3 4\pi P_0 \rho_0^2 (1 + x^2)^2)}\Bigg(\frac{\partial\ln P_0}{\partial r}  \chi (\gamma-1) (-\frac{\partial\ln P_0}{\partial r}  \nonumber\\ \nonumber
&& + (\frac{\partial\ln \rho_0}{\partial r} - \frac{\partial\ln \mu_0}{\partial r} +\frac{\partial\ln T_0}{\partial r}) \;\gamma\;)\; k_z^4 T_0 \Bigg ) \\ \nonumber
&& \Bigg (k_z^2\left(B_0^2 +4 \pi \rho_0 D^2 k_z^2\right)\left(\gamma D P_0 + \chi( \gamma - 1 ) T_0\right)\left(1+\frac{k_r^2}{k_z^2}\right)\\ \nonumber
&&- 4 \pi \left[\left(\frac{\partial P_0}{\partial r}\right)^2 D + \left(\frac{\partial P_0}{\partial r}\right) \left(-\frac{\partial\ln \rho_0}{\partial r} D \gamma P_0 + \chi \frac{\partial T_0}{\partial r} (\gamma-1) \right.\right.\\ \nonumber
&&\left. - \frac{\partial\ln \mu_0}{\partial r} \chi (\gamma - 1) T_0\right)-2\rho_0\Omega(\Omega + \Omega_r)(D \gamma P_0 + \chi ( \gamma - 1  ) T_0) \Bigg] \Bigg)\,,\\
\end{eqnarray}
\be
&&\Delta_4=\bigg[(\dfrac{\partial\ln P_0}{\partial r})^2 \frac{ \chi^2 ( \gamma-1)^3 k_z^8 T_0^2}{\gamma^4 16\pi^2 P_0^2 \rho_0^3 \left(1+\frac{k_r^2}{k_z^2}\right)^3}\bigg]\\\nonumber
&& \bigg[4\pi B_0^2(\gamma-1)\left[4\frac{\Omega^2}{k_z^2}+D^2k_z^2\left(1+\frac{k_r^2}{k_z^2}\right)\right]\left(\frac{\partial P_0}{\partial r}\right)^2\\\nonumber
&&+Dk_z^4B_0^4\left(D\gamma P_0+\chi(\gamma-1)T_0\right)\left(1+\frac{k_r^2}{k_z^2}\right)^2\\\nonumber
&&+D k_z^2 \left(1+\frac{k_r^2}{k_z^2}\right)\rho_0  (D \gamma P_0 + \chi (\gamma-1) T_0)  ( 4 \Omega^2 )(4\pi B_0^2 )\\\nonumber
&&+16\pi^2\left[\frac{\partial P_0}{\partial r}\left(\frac{\partial\ln \mu_0}{\partial r}-\frac{\partial\ln T_0}{\partial r}\right)+2\rho_0\Omega(\Omega+\Omega_r)\right]^2\\\nonumber
&&\times D\left(D\gamma P_0+\chi(\gamma-1)T_0\right)+4\pi\rho_0 D^3B_0^2k_z^6\left(1+\frac{k_r^2}{k_z^2}\right)^2\\\nonumber
&&\times\left(D\gamma P_0+\chi(\gamma-1)T_0\right)+ 16\pi^2\;(\gamma-1)D^2\left(\frac{\partial P_0}{\partial r}\right)^2\\\nonumber
&&\times \left[\frac{\partial P_0}{\partial r}\left(\frac{\partial\ln \mu_0}{\partial r}-\frac{\partial\ln T_0}{\partial r}\right)+2\rho_0\Omega(\Omega+\Omega_r)\right]\\\nonumber
&&+8\pi k_z^2D(B_0^2+2\pi\rho_0 D^2k_z^2)\left(D\gamma P_0+\chi(\gamma-1)T_0\right)\left(1+\frac{k_r^2}{k_z^2}\right)\\\nonumber
&&\times \left[\frac{\partial P_0}{\partial r}\left(\frac{\partial\ln \mu_0}{\partial r}-\frac{\partial\ln T_0}{\partial r}\right)+2\rho_0\Omega(\Omega+\Omega_r)\right] \bigg]
\en 
\begin{eqnarray}
&&\Delta_5=\bigg[(\dfrac{\partial\ln P_0}{\partial r})^4\dfrac{4 B_0^2 \chi^2 ( \gamma - 1)^4 k_z^{10} \Omega^2 T_0^2}{\gamma^5 64\pi^3 P_0 \rho_0^5 \left(1+\frac{k_r^2}{k_z^2}\right)^4}\bigg]\nonumber\\\nonumber
&&\bigg[ 4\pi(\gamma-1)DB_0^2\left(\frac{\partial P_0}{\partial r}\right)^2+(\gamma-1)\rho_0\,16\pi^2D^3k_z^2\left(\frac{\partial P_0}{\partial r}\right)^2\\\nonumber
&&+k_z^2B_0^4\left(D\gamma P_0+\chi(\gamma-1)T_0\right)\left(1+\frac{k_r^2}{k_z^2}\right)\\\nonumber
&&+8\pi\rho_0B_0^2D^2k_z^4\left(1+\frac{k_r^2}{k_z^2}\right)\left(D\gamma P_0+\chi(\gamma-1)T_0\right)\\\nonumber
&&+16\pi^2\rho_0^2D^4k_z^6\left(1+\frac{k_r^2}{k_z^2}\right)\left(D\gamma P_0+\chi(\gamma-1)T_0\right)\\\nonumber
&&+4\pi B_0^2\left[\frac{\partial P_0}{\partial r}\left(\frac{\partial\ln \mu_0}{\partial r}-\frac{\partial\ln T_0}{\partial r}\right)+2\rho_0\Omega(\Omega_r-\Omega)\right]\\\nonumber
&&\times \left(D\gamma P_0+\chi(\gamma-1)T_0\right)+\rho_016\pi^2D^2k_z^2\left(D\gamma P_0+\chi(\gamma-1)T_0\right)\\
&&\times \left[\frac{\partial P_0}{\partial r}\left(\frac{\partial\ln \mu_0}{\partial r}-\frac{\partial\ln T_0}{\partial r}\right)+2\rho_0\Omega(\Omega+\Omega_r)\right]\bigg]\,,
\end{eqnarray}
\begin{eqnarray}
\Delta_6 = \Delta_5\left[ 4 \pi \left\{\frac{d P_0}{dr}\left(-\frac{d\ln T_0}{dr} + \frac{d\ln \mu_0}{dr} \right)  + 2 \Omega r\frac{d\Omega}{dr} \rho_0\right\} \right.\nonumber \\ \left.+ B_0^2 k_z^2 \left(1+\frac{k_r^2}{k_z^2}\right)\right]\left[\frac{B_0^2 \chi D ( \gamma - 1) k_z^6 T_0 }{ 16 \pi^2 \gamma P_0 \rho_0^2 \left(1+\frac{k_r^2}{k_z^2}\right)}\right]\,.
\end{eqnarray}

\end{document}